\documentclass[10pt]{amsart}
 \title[Symmetry of stratified water waves]{Some criteria for the symmetry of stratified water waves}
 \date{\today}
 \author{Samuel Walsh}
 \address{Division of Applied Mathematics\\ 
Brown University \\ 
Providence, RI 02912 \\ 
USA.} 
\email{Samuel\_Walsh@brown.edu} 
\thanks{This work was supported in part by NSF grant DMS-0405066.} 
 
\usepackage{amsfonts, amsmath, amssymb, mathrsfs, txfonts}

\numberwithin{equation}{section}

\newcommand{\be}{\begin{equation} }
\newcommand{\ee}{\end{equation}}

\theoremstyle{plain} 
\newtheorem{theorem}{Theorem} 
\newtheorem{corollary}{Corollary} 
\newtheorem*{main}{Main Theorem} 
\newtheorem{lemma}{Lemma} 
\newtheorem{proposition}{Proposition} 
\theoremstyle{definition} 
\newtheorem{df}{Definition} 
\theoremstyle{remark} 

\newtheorem*{remark}{Remark}

\begin{document}

\begin{abstract}
This paper considers two-dimensional stably stratified steady periodic gravity water waves with surface profiles monotonic between crests and troughs.   We provide sufficient conditions under which such waves are necessarily symmetric.  This is done by first exploiting some elliptic structure in the governing equations to show that, in certain size regimes, a maximum principle holds. This then forms the basis for a method of moving planes argument.
\end{abstract}

\maketitle
\section{Introduction}
One of the characteristic features of ocean waves is their propensity to display stratification --- a heterogeneous distribution of density that can result from the interplay between gravity and the salinity of the water. In a typical scenario, one observes a region just below the surface where density increases quickly with depth, followed by a larger region extending to the ocean bed where it is essentially constant (cf. \cite{Pd}).  Qualitatively, flows with heterogeneous density may differ quite strikingly from their homogeneous counterparts.  For instance, slowly moving stratified traveling waves have a strong tendency to be two-dimensional.  That is, the motion is identical along any line perpendicular to the direction of propagation and the vertical axis \cite{Y1}.  Remarkably, this columnar behavior will persist even with a foreign body, such as an infinitely long cylinder, placed vertically within the fluid. 

Unsurprisingly, therefore, stratification has been the subject of a great deal of scholarly interest, especially in the oceanography and geophysical fluid dynamics communities.  In \cite{Wa, Wa2}, the author developed an existence theory for two-dimensional stratified steady and periodic gravity waves, with or without surface tension.  It was shown that, under certain unrestrictive assumptions, there is a global continuum of such waves that are classical.  This was done via a bifurcation argument, with the 1-parameter family of laminar flows playing the role of the trivial solutions.  As a consequence of this construction, it was shown that each member of the continuum has monotonic surface profile and is symmetric.  That is, the wave profile possesses a single crest and a single trough per period and is monotonic between them.  Moreover, both the horizontal velocity and surface are symmetric across the crest line, whereas the vertical velocity is antisymmetric.  In this work we address the complementary question: Under what circumstances are stratified steady water waves with monotonic profiles necessarily symmetric?  While at first blush it may seem that waves of this type are quite special, our main result implies that, in fact,  they are the only possible solutions within certain size regimes.

Let us now briefly review the setup for stratified waves in \cite{Wa}.  Fix a Cartesian coordinate system so that the $x$-axis points in the direction of propagation, and the $y$-axis is vertical.  We assume that the floor of the ocean is flat and occurs at $y = -d$.  Let $y = \eta(x,t)$ be the free surface.  We shall normalize $\eta$ by choosing the axes so that the free surface is oscillating around the line $y = 0$.  As usual we let $u = u(x,y,t)$ and $v = v(x,y,t)$ denote the horizontal and vertical velocities, respectively, and let $\rho = \rho(x,y,t) > 0$ be the density.

For water waves, it is appropriate to suppose that the flow is incompressible.  Mathematically, this assumption manifests itself as the requirement that the vector field be divergence-free for all time,
\be u_x + v_y = 0. \label{incompress} \ee
Taking the fluid to be inviscid, conservation of mass implies that the density of a fluid particle remains constant as it follows the flow.  This is expressed by the continuity equation
\be \rho_t + u \rho_x + v \rho_y = 0. \label{consmass1} \ee
Conservation of momentum is described by Euler's equations
\be \left \{ \begin{array}{lll}
\rho u_t + \rho u u_x + \rho v u_y  & = & -{P_x}, \\
\rho v_t + \rho u v_x + \rho v v_y & = & -{P_y} - g\rho, \end{array} \right. \label{euler1} \ee
where $P=P(x,y,t)$ denotes the pressure and $g$ is the gravitational constant.  Here we are assuming that gravity is the only external force acting on the fluid.  

On the free surface, the dynamic boundary condition requires the pressure of the fluid to match atmospheric pressure, that we shall denote $P_{\textrm{atm}}$.  Thus,
\be P = P_{\textrm{atm}}, \qquad \textrm{on } y = \eta(x,t). \label{presscond} \ee
The corresponding condition for the vector field is the so-called kinematic condition
\be v = \eta_t + u \eta_x, \qquad \textrm{on } y = \eta(x,t). \label{kincond1} \ee
In essence, \eqref{kincond1} states that fluid particles on the free surface remain there as the flow develops.  Finally, we assume that the ocean bed is impermeable and thus,
\be v = 0, \qquad \textrm{on } y = -d. \label{bedcond} \ee
Note that, since we are in the inviscid regime, we do not impose any conditions on $u$ along the bed. 
 
We are considering classical traveling periodic wave solutions $(u,v,\rho,P, \eta)$ to \eqref{incompress}--\eqref{bedcond}.  More precisely, we take this to mean that, for some fixed $c > 0$, the solution appears steady in time and periodic in the $x$-direction when observed in a frame that moves with constant speed $c$ to the right.  The vector field will thus take the form $u = u(x-ct,y)$,  $v = v(x-ct, y)$, where each of these is $L$-periodic in the first coordinate.  Likewise for the scalar quantities: $\rho = \rho(x-ct,y)$, $P = P(x-ct,y)$, and $\eta = \eta(x-ct)$, again with $L$-periodicity in the first coordinate.  We therefore take moving coordinates 
\[ (x-ct, y) \mapsto (x,y), \]
which eliminates time dependency from the problem.  In the moving frame \eqref{incompress}--\eqref{euler1} become 
\be \left \{ \begin{array}{lll} 
u_x + v_y & = & 0 \\
(u-c)\rho_x + v\rho_y & = & 0 \\
\rho (u-c)u_x + \rho v u_y  & = & -{P_x} \\
\rho (u-c) v_x + \rho v v_y & = & -{P_y} - g\rho \end{array} \right. \label{euler2} \ee
throughout the fluid domain.  Meanwhile, the reformulated boundary conditions are
\be \left \{ \begin{array}{llll} 
v & = & (u-c) \eta_x & \textrm{on } y = \eta(x) \\
v & = & 0 & \textrm{on } y = -d \\
P & = & P_{\textrm{atm}} & \textrm{on } y = \eta(x) \end{array} \right. \label{boundcond} \ee
where $(u,v,\rho, P)$ are taken to be functions of $x$ and $y$, $\eta$ is a function of $x$, and all of them are $L$-periodic in $x$.

Waves of this type will be called \emph{symmetric} provided that there exists $x_0 \in \mathbb{R}$ such that $(u,\rho, \eta)$ are symmetric over the line $x = x_0$, while $v$ is antisymmetric.  We shall call any local maximum of $\eta$ a \emph{crest}, and any local minimum a \emph{trough}.  

In the event that $u = c$ somewhere in the fluid we say that \emph{stagnation} has occurred, as in the moving frame the fluid appears to be stationary at that point.  As will become apparent later, stagnation points may produce some degeneracy or instability in the problem.  For that reason, we shall restrict our attention to the case where $u < c$ throughout the fluid.  

Recall that we have chosen our axes so that $\eta$ oscillates around the line $y = 0$.  In other words,
\be \fint_{-L/2}^{L/2} \eta(x)dx = 0. \label{normalsurface} \ee 
We shall also denote 
\[ \eta_{\max} := \max_{x \in \mathbb{R}} \eta(x) + d, \qquad \eta_{\min} := \min_{x \in \mathbb{R}} \eta(x) + d.\]
These are the maximum and minimum distances between the surface and the bed, respectively.  

Observe that, by conservation of mass and incompressibility, $\rho$ is transported and the vector field is divergence free.  Therefore we may introduce a (relative) \emph{pseudo-stream function} $\psi = \psi(x,y)$, defined uniquely up to a constant by: 
\[ \psi_x = -\sqrt{\rho}v,\qquad \psi_y = \sqrt{\rho} (u-c). \]
Here we have the addition of a $\rho$ term to the typical definition of the stream function for an incompressible fluid.  This neatly captures the inertial effects of the heterogeneity of the flow \cite{Y1}.  The particular choice of $\sqrt{\rho}$ is merely an algebraic nicety.

It is a straightforward calculation to check that $\psi$ is indeed a (relative) stream function in the usual sense, i.e. its gradient is orthogonal to the vector field in the moving frame at each point in the fluid domain.   As usual, we shall refer to the level sets of $\psi$ as the \emph{streamlines} of the flow.  Observe that, in assuming $u < c$, we have guarantee that the streamlines are not closed.  For definiteness we choose $\psi \equiv 0$ on the free boundary, so that $\psi \equiv p_0$ on $y = -d$, where $p_0$ is the (relative) \emph{pseudo-volumetric mass flux}, 
\be p_0 := \int_{-d}^{\eta(x)} \sqrt{\rho(x,y)}\left[u(x,y) - c\right] dy. \label{defp0} \ee
It is easy to check that $p_0$ is well-defined, i.e., the integral on the right-hand side above is independent of $x$.  Examining \eqref{defp0}, it is clear that $p_0$ describes the amount of fluid flowing through any vertical line extending from the bed to the free surface and with respect to the transformed vector field $(\sqrt{\rho}(u-c), \sqrt{\rho}v)$. 

Since $\rho$ is transported, it must be constant on the streamlines. Abusing notation we may therefore let $\rho \in  C^{2}([p_0,0] ; \mathbb{R}^{+})$ be given such that 
\be \rho(x,y) = \rho(-\psi(x,y)) \label{defrho} \ee
throughout the fluid.  When there is risk of confusion, we shall refer to the $\rho$ occurring on the right-hand side above as the \emph{streamline density function}.  We shall focus our attention on the case where the density is nondecreasing as depth increases.  This assumption is physically well-motived, since it is a prerequisite for hydrodynamic stability.  Note that the level set  $-\psi = p_0$ corresponds to the flat bed, and the set where $-\psi = 0$ corresponds to the free surface.  For that reason, we shall say the flow is \emph{stably stratified} provided that the streamline density function is nonincreasing. 

By Bernoulli's theorem, the quantity
\[ E := P + \frac{\rho}{2}\left( (u-c)^2 + v^2\right) + g\rho y, \]
is constant along streamlines.  Then, under the assumption that $u < c$ throughout the fluid, there exists a function $\beta \in C^1([0,|p_0|] ; \mathbb{R})$ such that 
\be \frac{dE}{d\psi}(x,y) = -\beta(\psi(x,y)). \label{defbeta} \ee
For want of a better name we shall refer to $\beta$ as the \emph{Bernoulli function} corresponding to the flow.  Physically it describes the variation of specific energy as a function of the streamlines.   It is worth noting that when $\rho$ is a constant, $\beta$ reduces to the vorticity function.

We now define several quantities describing various properties of the wave.  Let 
\[ D_\eta := \left\{ (x,y) \in \mathbb{R}^2 : -d < y < \eta(x) \right\}\]
denote the fluid domain (in the moving frame) and, for each $k \in \mathbb{N}$, domain $\Omega \subset \mathbb{R}^2$, we introduce the space 
\[ C^k_{\textrm{per}}(\overline{\Omega}) := \left\{ f \in C^k(\overline{\Omega}) : f(x + L,y) = f(x,y),~\forall x,y \in \overline{\Omega} \right\}.\]
Put 
\[ a_0 := \min_{\overline{D_\eta}} \sqrt{\rho} (c-u), \qquad A_0 := \max_{\overline{D_\eta}} \sqrt{\rho}(c-u).\]
The first of these gauges how far away from stagnation the flow is, while the second gives a bound on $u^-$, the motion of the fluid counter to the direction of propagation.  Next, we let 
\be M := \max \left\{\max_{\overline{D_\eta}} \left|\frac{v}{u-c}\right|,~ \max_{\overline{D_\eta}}  \left| \left(\partial_x + \frac{v}{u-c} \partial_y\right) \frac{v}{u-c} \right|, ~ \max_{\overline{D_\eta}} \left| \frac{1}{\sqrt{\rho}(u-c)} \partial_y \frac{v}{u-c} \right| \right\}. \label{defM} \ee
$M$ describes  (in a sense that will be made clear later) the variation of $dy/dx$ following along a streamline.  In particular, $M = 0$ only for the family of laminar flows, these being shear flows where the streamlines are parallel to the bed and $\eta \equiv 0$.  The motivation here will become significantly more transparent in the next section, in particular, see \eqref{defM2} and the accompanying discussion.

Our main result is the following.  
\begin{main} Consider a stably stratified steady periodic wave train propagating at fixed speed $c$ over a flat bed at $y = -d$ with relative pseudo-mass flux $p_0 < 0$.  Let $\rho \in C^2([p_0, 0]; \mathbb{R}^+)$ and $\beta \in C^1([0, |p_0|])$ be the streamline density function and Bernoulli function associated with the flow, and let $(u,v) \in C^2_{\textrm{per}}(\overline{D_\eta}) \times C^2_{\textrm{per}}(\overline{D_\eta})$ be the vector field.  Assume that the wave profile $y = \eta(x)$ is monotonic between crests and troughs with period $L$ and that $u-c < 0$.    Each of the following is a sufficient condition for the wave to be symmetric.
\be \eta_{\max}^2 \sup{\beta^\prime}  + g \eta_{\max}^3 \sup{(\rho^{\prime\prime})^+} < \pi^2, \qquad 0 \leq s \leq |p_0|; \tag{S1} \label{theorem1cond1}  \ee
or
\be \sup \beta^\prime + g \eta_{\max} \sup{(\rho^{\prime\prime})^+} < \exp{ \left(-\min\{L,~ \eta_{\min} \} \right)}; \tag{S2} \label{theorem1cond2} \ee
or 
\be \left\{ \begin{array}{c}
A_0 \epsilon_1 \sqrt{\frac{\epsilon_2}{a_0}} + \epsilon_2 \sqrt{\frac{\epsilon_2}{a_0}} + g \sup \left|\rho^\prime \right| < \exp{\left({-a_0^{-1/2} \min\{L, |p_0|\}}\right)},  \\
\epsilon_1 < \frac{a_0^2}{A_0}, \qquad \epsilon_2 < a_0, \end{array} \right.
 \tag{S3} \label{theorem1cond3} \ee
 where
\[ \epsilon_1 := 3g \eta_{\max} \sup \left| \rho^\prime \right|, \]
and
\begin{align*} \epsilon_2 & :=  4MA_0^2 +2M^2 A_0^3 + 3A_0 \left( \sup |\beta| + g \eta_{\max} \sup |\rho^\prime| \right) \nonumber \\ 
&  \qquad + 2M^3 a_0^{-1} A_0^3 + M^2 a_0^{-1} A_0^3 + M a_0^{-3} A_0^3\left(\sup|\beta| + g\eta_{\max} \sup |\rho^\prime| \right). \end{align*}
\label{theorem1}
\end{main}

\begin{remark} \label{maintheoremremark} The first condition \eqref{theorem1cond1} is a direct generalization of the main result of \cite{CE},  while the second follows from a similar argument but takes a slightly different view.  We mention  that, in particular, the condition $\beta^\prime,~\rho^{\prime\prime} \leq 0$ immediately implies both of conditions \eqref{theorem1cond1}--\eqref{theorem1cond2}.  Note that this is a physically natural assumption on $\rho$ for ocean waves \cite{Pd}.  For $\beta^\prime$ or $\rho^{\prime\prime}$ small but positive, the  {Main Theorem} implies that the wave is symmetric provided that the maximum elevation of the profile and/or the period are sufficiently small.  

On the other hand, condition \eqref{theorem1cond3} gives a criteria based on the distance from stagnation, the degree to which the flow is laminar, as well the magnitudes of the volumetric mass flux, Bernoulli function, density variation and the period.  Generally, it states that solutions exhibiting small $\beta$ and $\rho^\prime$ and which are sufficiently far from stagnant are necessarily symmetric.    \end{remark}

Let us now discuss the history of this problem.  The basic machinery we employ is the moving planes method, which was originated by Alexandroff \cite{Al}.  Since then, this argument has been used extensively by a number of authors, notably Gidas, Nirenberg and Ni \cite{GNN} and Serrin \cite{Sr}.  The first significant application of a moving planes argument to the study of water waves was by Craig and Sternberg \cite{CrSt}, who considered the case of solitary waves in the irrotational setting.

The direct inspiration for our work is a number of recent breakthroughs in the study of symmetry for two-dimensional constant density rotational gravity water waves, beginning with a result of Constantin and Escher \cite{CE} analogous to condition \eqref{theorem1cond1} of our main theorem.  This was a generalization of an earlier result of Okamoto and Sh\=oji, who showed in \cite{OS} that symmetry must occur when the profile has a single crest and is monotonically decreasing.  Soon thereafter, Hur, building on the work of Garabedian \cite{Gn} and Toland \cite{Tl} for irrotational waves, proved symmetry for arbitrary vorticity functions under the assumption that every streamline attains a minimum on the trough line \cite{Hr}.  While this last condition was not entirely physical, she also introduced the idea of working within the Dubreil--Jacotin formulation, which proved to be extremely convenient.  As we shall see, the main advantage of Dubreil--Jacotin is that it demotes the vorticity function from a semilinear forcing term to the coefficient of a first-order term in a quasilinear equation, thereby nullifying its importance for symmetry.  Refining Hur's argument,  Constantin,  Ehrnstr\"om and Wahl\'en were able to significantly improve upon the result in \cite{CE}.  In \cite{CEW}, they showed that \emph{every} two-dimensional rotational steady gravity wave with a monotonic profile is symmetric.  In other words, symmetry follows without any assumptions on the vorticity function.

Work on the heterogeneous case has been comparatively sparse.  The most complete results are due to Maia \cite{Ma}, who considered solitary internal waves confined to an infinite horizontal strip.  It was assumed that the waves were quiescent at $+\infty$ and connected to a laminar flow at $-\infty$, where a piece-wise continuous density distribution was prescribed.  Maia proved that, in this setting, waves of elevation are symmetric.  Her argument, however, relies heavily on both the geometry of the domain and the type of boundary data at $\pm\infty$, and therefore does not extend to the free surface and periodic case which we consider.

The structure of our argument is as follows.  In \S\ref{reformulation} we begin by providing two scalar reformulations of \eqref{euler2}--\eqref{boundcond}; these are the natural generalizations to heterogeneous flows of the equations studied in \cite{CEW} and \cite{CE}.  In \S\ref{maximumprinciple}, we then derive estimates on the given quantities that ensure solutions of these equations satisfy a maximum principle.  This sets the stage for the moving planes argument that we provide in \S\ref{symmetry}.

\section{Preliminaries} \label{preliminaries}

\subsection{Two reformulations} \label{reformulation}

It will be convenient to replace the system \eqref{euler2}-\eqref{boundcond} with an equivalent scalar PDE for the stream function.  Differentiating the expression for $E$ and appealing to the definition of $\psi$, one can prove the following identity, known as Yih's equation or the Yih--Long equation \cite{Y1}, 
\[ \frac{d E}{d \psi} = \Delta \psi + gy \frac{d \rho}{d \psi}. \]
Rewriting the above expression in terms of the Bernoulli function $\beta$ and streamline density function yields the following
\be -\beta(\psi) = \Delta \psi - gy \rho^\prime(-\psi). \label{yiheq} \ee 
Moreover, evaluating Bernoulli's theorem on the free surface $\psi \equiv 0$, we find
\be |\nabla \psi|^2 + 2g \rho(-\psi) \left(\eta(x)+d\right) = Q, \qquad \textrm{on } y = \eta(x) \label{defQ} \ee
where the constant $Q := 2(E|_\eta-P_{\textrm{atm}} + gd)$.  Note that $Q$ gives roughly the energy density along the free surface of the fluid.  Together with the fact that $\psi = -p_0$ on $y = -d$ and $\psi = 0$ on the free surface, \eqref{yiheq}--\eqref{defQ} provide a complete reformulation of the problem as a semilinear elliptic equation.  

It will also prove useful to consider the semi-Lagrangian (or Dubreil--Jacotin) formulation, wherein we trade some additional nonlinearity in exchange for a fixed the domain.  Consider the alternate coordinate system, $(q,p)$, where 
\[ q = x, \qquad p = -\psi(x,y).\]
Then, under the transformation $(x,y) \mapsto (q,p),$ the closed fluid domain is mapped to the rectangle
\[ \overline{R} := \{ (q,p) \in \mathbb{R}^2 : -L/2 \leq q \leq L/2,~p_0 \leq p \leq 0 \}.\]
The purpose of the minus sign is simply to flip the rectangle so that the free surface will correspond to the top of $R$, while the flat bed will mapped to the bottom.  Given this, we shall denote
\[ T := \{ (q,p) \in R : p = 0 \}, \qquad B := \{ (q,p) \in R : p = p_0 \}. \]
  
Note that, in light of \eqref{defrho} and \eqref{defbeta}, we have that
\[ \beta = \beta(-p), \qquad \rho = \rho(p).\]
Moreover, the assumption that the streamline density function is nonincreasing becomes
\be \rho_p \leq 0. \label{rhoincdepth} \ee
  
Next, following the ideas of Dubreil--Jacotin, define 
\be h(q,p) := y + d \label{defh} \ee
which gives the height above the flat bottom on the streamline corresponding to $p$ and at $x = q$.  We calculate:
\be \psi_y = -\frac{1}{h_p}, \qquad \psi_x = \frac{h_q}{h_p} \label{hqhpeq}, \ee
and
\be \partial_q = \partial_x + h_q \partial_y, \qquad \partial_p = h_p \partial_y. \label{chainrule} \ee
This is the motivation for the definition of $M$ in \eqref{defM}.  Using \eqref{hqhpeq} to write \eqref{chainrule} in terms of the Eulerian quantities $(u,v)$, we see that 
\be M = \| h_{q} \|_{C^1(\overline{R})}. \label{defM2} \ee
For a laminar flow, of course, we have $h_q, h_{qq}, h_{qp} \equiv 0$.  It is in this sense that $M$ describes the distance of the flow from laminar. 

Observe that \eqref{hqhpeq} also implies $h_p > 0$, because we have stipulated that $u < c$ throughout the fluid.  Yih's equation \eqref{yiheq} becomes the following
\be -h_p^3 \beta(-p) =  (1+h_q^2)h_{pp} + h_{qq}h_p^2 - 2h_q h_p h_{pq} - g(h-d) h_p^3 \rho_p \label{interheighteq} \ee
where we have used \eqref{defh} to write $y = h-d$.  Recall, however, that we have normalized $\eta$ so that it has mean zero.  Taking the mean of \eqref{defh} along $T$, we obtain
\be d = d(h) = \fint_{-L/2}^{L/2} h(q,0)dq. \label{defd} \ee
That is, the average depth $d$ must be viewed as a linear operator acting on $h$.  Namely, it is the average value of $h$ over $T$.  Where there is no risk of confusion, we shall suppress this dependency and simply write $d$.  

Next consider the boundaries of the transformed domain.  On the bed we must have, by the definition of $h$, that 
\be h \equiv 0, \qquad \textrm{on } B. \label{heightboundcondB} \ee
In the new coordinates, the definition of $Q$ given in $\eqref{defQ}$ becomes the requirement
\be 1 + h_q^2+ h_p^2\left( 2g \rho h - Q\right) = 0, \qquad \textrm{on } T. \label{heightboundcondT} \ee

Altogether, then, combining \eqref{interheighteq}, \eqref{heightboundcondB} and \eqref{heightboundcondT} we have that the fully reformulated problem is the following. Find $(h,~Q) \in C^{3}_{\textrm{per}}(\overline{R}) \times \mathbb{R}$ satisfying
\be \left \{ \begin{array}{lll}
(1+h_q^2)h_{pp} + h_{qq}h_p^2 - 2h_q h_p h_{pq} - g(h-d(h))h_p^3 \rho_p = -h_p^3 \beta(-p) & p_0 < p < 0 \\
1+h_q^2 + h_p^2( 2g \rho h - Q)  = 0 & p = 0, \\
h = 0 & p = p_0, \end{array} \right. \label{heighteq} \ee  \\
where $h_p > 0$.  Here $\rho \in C^{2}([p_0,0];\mathbb{R}^+)$ and $\beta \in C^{1}([0,|p_0|])$ are given function with $\rho_p \leq 0$.

\subsection{Maximum principle} \label{maximumprinciple}
In anticipation of our argument in the next section, we now attempt to find an elliptic problem satisfied by the difference of two solutions of the height equation.  Fix $h,\tilde{h}$ solutions to \eqref{heighteq} with the same value of $Q$ and suppose that $d(h) = d(\tilde{h})$.  Define
\be \begin{array}{lll} \mathcal{L} & := &
(1+h_q^2)h_p^{-3} \partial_p^2 + h_p^{-1} \partial_q^2 - 2 h_q h_p^{-2} \partial_p \partial_q \\ 
& & + \left(\tilde{h}_{qq}(h_p+\tilde{h}_p) - 2\tilde{h}_q \tilde{h}_{pq} + \beta(-p)(h_p^2 + h_p \tilde{h}_p + \tilde{h}_p^2)\right) h_p^{-3} \partial_p \\
& & + \left(\tilde{h}_{pp} (h_q + \tilde{h}_q) -2h_p \tilde{h}_{pq}\right) h_p^{-3} \partial_q \\
& & -  g\rho_p (\tilde{h}-d(\tilde{h}))(h_p^2+h_p \tilde{h}_p + \tilde{h}_p^2) h_p^{-3}\partial_p \\ 
& & -g\rho_p . \end{array} \label{defL} \ee 
Note that the first three lines of \eqref{defL} are the same as the analogous operator found in \cite{CEW}, except that we have divided through by $h_p^3$;  the stratification manifets itself only in the last two lines.  Also, since $\rho_p \leq 0$, the sign of the zeroth-order term ``goes the wrong way'' in the sense that, even taking for granted that $\mathcal{L}$ is elliptic, the maximum principle does not generally apply.  The goal of this section is to find suitable restrictions on $h$ so that a maximum principle like result \emph{will} hold.  Our main tool for this is a number of results due to Berestycki, Nirenberg and Varadhan \cite{BNV}.  We begin by summarizing the most relevant of these.

\begin{df}  Let $\mathcal{L}$ be an elliptic second-order linear differential operator with bounded continuous coefficients.  The \emph{principal eigenvalue} $\lambda_1 = \lambda_1(\mathcal{L})$ is defined as follows
\[ \lambda_1 := \sup \{ \lambda \in \mathbb{R} : \exists \phi > 0 \textrm{ in } R \textrm{ satisfying } (\mathcal{L}+\lambda) \phi \leq 0 \}. \] \label{deflambda1} \end{df}
The fact that $\lambda_1$ is well-defined, i.e., that the supremum above exists, is a classical result.  Our interest in $\lambda_1$ stems from the following theorem.

\begin{theorem} \emph{({BNV}, \cite{BNV})} Let $\mathcal{L}$ be an elliptic second-order linear differential operator with bounded continuous coefficients.  Then $\mathcal{L}$ has a (refined) maximal principle if and only if $\lambda_1(\mathcal{L}) > 0$.  \label{bnvtheorem1} \end{theorem}
The meaning of ``refined'' above is made precise in \cite{BNV}, but for our purposes it can be safely said to coincide with the traditional definition of the (weak) maximum principle.  

In light of  {Theorem \ref{bnvtheorem1}}, our objective is clear: We seek to prove lower bounds on the principal eigenvalue of the operator in \eqref{defL}.  The most basic such estimate is given by the following result of Protter and Weinberger \cite{PW}.

\begin{theorem} \emph{( {PW}, \cite{PW})} Let $\mathcal{L}$ be a general second order linear differential operator with bounded coefficients.  Then, if $\phi \in C^2(R) \cap C^1(\overline{R})$ is a strictly positive function on $R$, 
\[ \lambda_1 (\mathcal{L}) \geq \inf_{R} \left(-\frac{\mathcal{L}\phi}{\phi}\right).\]
\label{pwtheorem}
\end{theorem}  
Using this with together with a construction in \cite{BNV} we can prove a very rough but explicit bound on $\lambda_1$ in the case where there is no zeroth-order term.

\begin{corollary} Let $M := a_{ij} \partial_i \partial_j + b_i \partial_i$ be a second-order elliptic linear differential operator with bounded continuous coefficients.  Let $a_0$ denote the lower ellipticity constant and put $b^2 := \sum b_i^2$.  Let $\sigma > 0$ solve the equation
\[ a_0 \sigma^2 - b\sigma - b = 1.\]
Then
\be \lambda_1(M) \geq \exp{ \left(-\sigma \min\{L, |p_0|\}\right)}.\label{Mlowerbound} \ee
\label{boundonlambda1lemma}
\end{corollary}
\begin{proof} In \cite{BNV} it was shown that there exists a function $u_0$ satisfying $M u_0 = -1$ in $R$, $u > 0$ in $R$ and $u|_{\partial R} = 0$ (potentially in some weak sense.)  If we further examines the actual construction of $u_0$, we find that $0 < u_0 \leq e^{\sigma \min\{L, |p_0|\}}$, with $\sigma$ as above (see \S3 of \cite{BNV}, particularly p. 61).    Then \eqref{Mlowerbound} follows directly from  {Theorem \ref{pwtheorem}} with $\phi = u_0$.   \end{proof}

Finally, for reference we repackage two important estimates from \cite{BNV} regarding the dependence of $\lambda_1$ on the magnitudes of the first- and zeroth-order terms.

\begin{proposition} \emph{( {BNV})} Let $M := a_{ij} \partial_i \partial_j + b_i \partial_i$ and $M^\prime :=  a_{ij} \partial_i \partial_j + b_i^\prime \partial_i$ be a second-order elliptic differential operators with smooth coefficients. Let $a_0$ be the lower ellipticity constant and put $b^2 := \sum_i b_i^2$.  Suppose that $\delta$ is chosen so that
\[ \sum_i \left(b_i^\prime - b_i\right)^2 \leq \delta^2 \leq b a_0,\] 
then
\be \lambda_1(M^\prime) \geq \lambda_1(M) - \sqrt{\frac{b}{a_0}} \delta. \label{MMprimeestimate} \ee
\label{bnvtheorem2}
\end{proposition}
\begin{proof} This is essentially estimate (5.2) of \cite{BNV} (used to prove  {Proposition 5.1} of that paper) with $c \equiv 0$.  \end{proof}

\begin{proposition} \emph{( {BNV})} In its dependence on the zeroth-order term, $\lambda_1$ is Lipschitz continuous in the $L^\infty$-norm with Lipschitz constant 1. \label{bnvtheorem3} \end{proposition}

That said, we are now prepared to prove the key lemma of this section.
\begin{lemma}
\emph{(Properties of $\mathcal{L}$)} Let $h,~\tilde{h} \in C_{\mathrm{per}}^{3+\alpha}(\overline{R})$, $Q \in \mathbb{R}$ be given such that $d(h) = d(\tilde{h})$ and $(h,Q),~(\tilde{h},Q)$ are solutions of \eqref{heighteq} with $\sup_{\overline{R}} h_p^{-1} = \sup_{\overline{R}} \tilde{h}_p^{-1}$, $\inf_{\overline{R}} h_p^{-1} = \inf_{\overline{R}} \tilde{h}_p^{-1}$ .  Then the operator $\mathcal{L}$ defined in \eqref{defL} satisfies the following.  
\begin{itemize}
\item[\emph{(a)}] $\mathcal{L}(h-\tilde{h}) = 0$.  
\item[\emph{(b)}] $\mathcal{L}$ is a uniformly elliptic differential operator with lower ellipticity constant $a_0 = \inf_{\overline{R}} h_p^{-1}$.
\item[\emph{(c)}] Let $\lambda_1(\mathcal{L})$ denote the principle eigenvalue of $\mathcal{L}$ as in \eqref{deflambda1}.  Define
\be \begin{split}
b &:=  \sup_{\overline{R}} \bigg\{ \Big[ h_p^{-3}\tilde{h}_{qq}(h_p+\tilde{h}_p) - 2h_p^{-3}\tilde{h}_q \tilde{h}_{pq} + \beta(-p)h_p^{-3}(h_p^2 + h_p \tilde{h}_p + \tilde{h}_p^2) \\ 
& \qquad -h_p^{-3}g\rho_p(\tilde{h}-d(\tilde{h}))(h_p^2 + h_p \tilde{h}_p + \tilde{h}_p^2) \Big]^2 \\  
& \qquad + \left[ h_p^{-3}\tilde{h}_{pp} (h_q + \tilde{h}_q) -2h_p^{-2} \tilde{h}_{pq}\right]^2 \bigg\}^{\frac{1}{2}}. 
\end{split} \label{defb} \ee
and let $\sigma$ be as in  {Corollary \ref{boundonlambda1lemma}}.  Then $\lambda_1(\mathcal{L})$ is strictly positive provided that
\be e^{-\sigma \max\{L,~|p_0|\}} > \sup_{[p_0, 0]} |g \rho_p |. \label{condition1} \ee
Alternatively, if we recall $A_0 = \sup_{\overline{R}} h_p^{-1}$ and define the constants
\be \delta_1 := 3|p_0| \sup_{[p_0, 0]} |g\rho_p|, \label{defdelta1} \ee
and
\be \begin{split} 
\delta_2 &:= \sup_{\overline{R}} \bigg\{ \left[ h_p^{-3}\tilde{h}_{qq}(h_p+\tilde{h}_p) - 2h_p^{-3}\tilde{h}_q \tilde{h}_{pq} + \beta(-p)h_p^{-3}(h_p^2 + h_p \tilde{h}_p + \tilde{h}_p^2) \right]^2 \\  
& \qquad + \left[ h_p^{-3}\tilde{h}_{pp} (h_q + \tilde{h}_q) -2h_p^{-2} \tilde{h}_{pq}\right]^2 \bigg\}^{\frac{1}{2}}, 
\end{split} \label{defdelta2} \ee
then $\lambda_1(\mathcal{L})$ is strictly positive provided that
\be \delta_1 < \frac{a_0^2}{A_0}, \qquad \delta_2 < a_0, \label{eigenvaluecondition1} \ee
and
\be  A_0 \delta_1 \sqrt{\frac{b}{a_0}} + \delta_2 \sqrt{\frac{\delta_2}{a_0}} + \sup_{[p_0,0]} |g\rho_p| < \exp{\left(-a_0^{-1/2} \min\{L,~|p_0|\} \right)} . \label{eigenvaluecondition2} \ee
\end{itemize}
\label{propLlemma} \end{lemma}
\begin{proof}  A tedious but easy calculation readily confirms that $\mathcal{L}(h-\tilde{h}) = 0$, proving (a).   Similarly, the second-order terms of $\mathcal{L}$ are identical to those of the differential operator in the height equation which has been proven to be uniformly elliptic \cite{Wa}.   Calculating the precise lower ellipticity constant is a simple exercise that we omit for brevity.  

The interesting part is, of course, proving (c), which is done as follows.  By  {Proposition \ref{bnvtheorem3}} we know that $\lambda_1$ is Lipschitz in its dependence on the zeroth-order coefficients with Lipschitz constant 1.  That is, if we denote as $\mathcal{L}_0$ the differential operator found by dropping the zeroth-order term in $\mathcal{L}$, then 
\be \lambda_1(\mathcal{L}) > \lambda_1(\mathcal{L}_0) - \sup |g \rho_p|. \label{LL0estimate} \ee
In order to show $\lambda_1(\mathcal{L})$ is positive, therefore, we must produce lower bounds for $\mathcal{L}_0$.  This can be achieved by using  {Corollary \ref{boundonlambda1lemma}} directly, proving $\lambda_1(\mathcal{L})$ is positive under condition \eqref{condition1}. 

To arrive at the more involved conditions \eqref{eigenvaluecondition1}-\eqref{eigenvaluecondition2} we instead begin by chopping off the first-order terms involving $\rho_p$ using  {Proposition \ref{bnvtheorem2}}.  Let $\mathcal{L}_1$ denote the operator found by setting $\rho_p \equiv 0$ in $\mathcal{L}$.  Let $b$ be as in \eqref{defb} and put
\[ \delta := \sup_{\overline{R}} \left| g\rho_p h_p^{-3}(\tilde{h}-d(\tilde{h}))(h_p^2+h_p \tilde{h}_p + \tilde{h}_p^2)\right|.\]
Under the assumption that $\delta^2 < b a_0$ we have by \eqref{MMprimeestimate}:
\be \lambda_1(\mathcal{L}_0) \geq \lambda_1(\mathcal{L}_1) - \sqrt{\frac{b}{a_0}} \delta. \label{L0L1estimate}\ee
So that we can better understand the above statement we now endeavor to control $\delta$.  We can bound $\tilde{h} - d(\tilde{h})$ by $\tilde{\eta}+d$, where $\tilde{\eta}$ is the free surface corresponding to $\tilde{h}$.  Since we have stipulated that $\sup_{\overline{R}} h_p = \sup_{\overline{R}} \tilde{h}_p$,
\[ \delta \leq  3 A_0 \sup_{[p_0, 0]} | g \rho_p | ~ \sup_{T} \left| \tilde{\eta} +d \right|.\]
As $\delta < b$, the assumption $\delta^2 < ba_0$ can be satisfied by requiring that $\delta < a_0$, or equivalently
\[ 3 \sup_{[p_0, 0]} |g \rho_p | ~\sup_T | \tilde{\eta}+d| < a_0/A_0. \]
We remark that because $a_0 \leq A_0$, the left-hand side above is necessarily less than one.  It warrants notice that this implies there is a definite bound on either the magnitude of the density variation, or alternatively on the amplitude of the wave.  We can eliminate $\tilde{\eta}+d$ if we are willing to further estimate $\tilde{\eta}+d(\tilde{h}) \leq |p_0| \sup_{\overline{R}} h_p =  |p_0| a_0^{-1}$.  Plugging this into the above inequality reveals that it is sufficient to take $\delta_1 = 3|p_0| \sup |g\rho_p|  < a_0^2/A_0$ in order to ensure estimate \eqref{L0L1estimate} holds.

Using this information in concert with \eqref{LL0estimate} we obtain
\begin{align*} \lambda_1(\mathcal{L}) &> \lambda_1(\mathcal{L}_1) -  \sup |g \rho_p| - \sqrt{\frac{b}{a_0}} \delta \\ 
& >  \lambda_1(\mathcal{L}_1) -  \sup |g \rho_p| -  A_0 \delta_1 \sqrt{\frac{b}{a_0}}. \end{align*}

Next, let us denote by $\mathcal{L}_2$ the operator consisting of the second-order terms of $\mathcal{L}$.  Applying  {Corollary \ref{boundonlambda1lemma}} to $\mathcal{L}_2$ yields
\[ \lambda_1(\mathcal{L}_2) \geq e^{-a_0^{-1/2} \min\{L,~|p_0|\}}.\]
The final step is to estimate the difference between $\lambda_1(\mathcal{L}_1)$ and $\lambda_1(\mathcal{L}_2)$.  By the same argument as \eqref{L0L1estimate}  we get
\be \lambda_1(\mathcal{L}_1) \geq \lambda_1(\mathcal{L}_2) - \sqrt{ \frac{\delta_2}{a_0}} \delta_2, \label{L1L2estimate} \ee
since we have $\delta_2 < a_0$ by \eqref{eigenvaluecondition1}.

Incorporating the last estimate with \eqref{LL0estimate}-\eqref{L0L1estimate}, yields the condition in \eqref{eigenvaluecondition2}.  Our two assumptions on the relative size of $a_0$ and $(\delta_1,~\delta_2)$ --- made when deriving \eqref{L0L1estimate} and \eqref{L1L2estimate} --- are collected in \eqref{eigenvaluecondition1}.  This completes the proof of (c) and the lemma. \end{proof}

\section{Symmetry} \label{symmetry}

We are now prepared to prove the main theorem.  This shall be done by employing an adapted moving plane method (cf. \cite{CEW, CE, GNN, Sr}.)  The necessary maximum principle properties will follow from the results of the previous section.

\begin{proof}[Proof of Main Theorem] First consider the case when \eqref{theorem1cond1} holds.  We shall structure our argument on those of \cite{CE}.  Without loss of generality, suppose that the trough of the wave train occurs at $x = \pm L/2$.  Also, since we will use the Euler formulation exclusively for \eqref{theorem1cond1}--\eqref{theorem1cond2}, we shall make a change of variables so that the bed occurs at $y = 0$ and the free surface is $y = \eta(x) + d$.  This has the convenient effect of making $y$ non-negative. 

By assumption, we may choose $\delta \in (0,1/2)$ satisfying
\be \beta^\prime(s) \eta_{\max}^2 + g\rho^{\prime\prime}(-s) \eta_{\max}^3 \leq \pi^2\left(1-2\delta\right)^2, \qquad 0 \leq s \leq |p_0|,  \ee
We may therefore define a function $\alpha \in C^2([0,\eta_{\max}])$ by
\[ \alpha(y) := \sin{ \pi\left( (1-2\delta) \frac{y}{\eta_{\max}} + \delta \right)}, \qquad 0 \leq y \leq \eta_{\max}. \]
One can readily verify that $\alpha$ has the following two properties 
\be 0 < \sin{(\pi \delta)} \leq \alpha \leq 1, \qquad  \frac{\alpha_{yy}}{\alpha} = -\frac{\pi^2(1-2\delta)^2}{\eta_{\max}^2} \geq - \frac{\pi^2}{\eta_{\max}^2}.\label{propalpha} \ee   

For each $\lambda \in (-L/2, 0]$ consider the truncated fluid domain 
\[ D_{\lambda} := D \cap \{ (x,y) : -L/2 < x < \lambda \}. \]
Since the surface profile is monotonically decreasing with the trough occurring at $x = -L/2$, for $\lambda$ sufficiently near $-L/2$, the transformation $(x,y) \mapsto (2\lambda-x, y)$ maps $D_\lambda$ into $D$.  For such $\lambda$ we define the reflected domain 
\[ D_\lambda^{R} := D_\lambda \cup \{ (2\lambda - x,y) : (x,y) \in D_\lambda \} \subset D. \] 
Incrementing $\lambda$, we eventually reach a point where $D_\lambda^R$ is no longer contained in $D$.  Therefore, if we put 
\[ \lambda_0 := \inf~\left\{ \lambda \in (-L/2,0] : D_\lambda^R \subset D \right\}, \]
We consider three possibilities, the first being that $\lambda_0 = 0$.  If this fails to be true, then $\lambda_0 < 0$ and we further consider the cases where $\lambda_0$ occurs at a crest, and where $\lambda_0$ occurs strictly to the left of a crest.   \\

\noindent \emph{Case 1}:  $\lambda_0 = 0$.  For each $(x,y) \in D_{\lambda_0}^R =: \Omega$, put
\[ w(x,y) := \frac{\psi(-x,y) - \psi(x,y)}{\alpha(y)},\]
where $\psi$ is the pseudo-stream function.  Proving symmetry of the wave is equivalent to showing that $w \equiv 0$ in $\Omega$.  Note that $\psi(-x,y) \geq \psi(x,y)$ on the free surface, by construction, and $\psi(-x,y) = \psi(x,y)$ on the bed and under the trough line, by periodicity.   In other words, $w \geq 0$ on $\partial \Omega$.  This suggests that, in order to prove $w$ vanishes identically,  we need to show that $w$ satisfies a maximum principle.   Naturally, this will have to come from \eqref{yiheq}.

With that in mind, we compute
\[ \Delta w + 2\frac{\alpha_y}{\alpha} \partial_y w = \frac{\Delta (\alpha w)}{\alpha} - \frac{\alpha_{yy}}{\alpha} w, \qquad \textrm{in } \Omega.  \]
Since $\alpha(y) w(x,y) = \psi(-x,y) - \psi(x,y)$, from \eqref{yiheq} and \eqref{propalpha} we conclude
\[ \Delta w + 2\frac{\alpha_y}{\alpha} \partial_y w + c w \leq 0, \qquad \textrm{in } \Omega,\]
where  
\[ c :={\overline{\beta}}{} - gy{ \overline{\rho^\prime}}{} - \frac{\pi^2}{\eta_{\max}^2},\]
and, for all $(x,y) \in \Omega$,
\[ \overline{\beta}(x,y) := \beta(\psi(-x,y)) - \beta(\psi(x,y)), \qquad \overline{\rho^\prime}(x,y) := \rho^\prime(-\psi(-x,y)) - \rho^\prime(-\psi(x,y)).\]  
But, by the mean value theorem, for each $(x,y) \in \Omega$, there exist $s_1=s_1(x,y) \in (0,|p_0|)$, $s_2=s_2(x,y) \in (p_0,0)$, such that $\overline{\beta}(x,y) = \beta^\prime(s_1)$, $\overline{\rho^\prime}(x,y) = -\rho^{\prime\prime}(s_2)$.  Since $0 \leq y \leq \eta_{\max}$, we have
\[ c(x,y) \leq \beta^\prime(s_1) + g\eta_{\max} \left(\rho^{\prime\prime}(s_2)\right)^+ - \frac{\pi^2}{\eta_{\max}^2}, \qquad \forall (x,y) \in \Omega \] 
which is non-positive by condition \eqref{theorem1cond1}.  We are therefore able to apply a minimum principle and conclude either $w > 0$, or  $w \equiv 0$ in $\Omega$.

Let us denote the corner point $C := (-L/2, \eta(-L/2))$. Then, by periodicity of the wave profile, we must have
\[w(C) = w_y(C) = w_{yy}(C) = w_{xx}(C) = 0.\]  
Likewise, differentiating the relation $\psi = 0$ on$y = \eta(x)$, we find 
\[ \psi_x+\psi_y \eta^\prime = 0, \qquad \textrm{on } y = \eta(x). \]  
Since the trough occurs at $x = -L/2$, we have $\eta^\prime(-L/2) = \psi_x(C) = 0$.   It follows that $w_x(C) = 0$.  

Finally, consider $w_{xy}(C)$.  From the nonlinear boundary condition \eqref{defQ} we compute
\[ \psi_x (\psi_{yy} + \psi_{xy} \eta^\prime) + \psi_y (\psi_{xy} + \psi_{yy} \eta^\prime) - g\rho^\prime(-\psi) \psi_x (\eta(x)+d) + g\rho(-\psi) \eta^\prime = 0, \qquad \textrm{on } y = \eta(x).\]
Evaluating this at $C$, we conclude $\psi_{xy}(C) \psi_y(C) = 0$.  But, since $\psi_y < 0$, this can only be the case provided $\psi_{xy}$ vanishes at $C$.  In light of this, and the fact that $w_x(C) = 0$, it follows that 
\[ w_{xy} = -2 \frac{\psi_{xy}}{\alpha} -\frac{\alpha_y}{\alpha} w_x = 0, \qquad \textrm{at } C.\]
Thus all derivatives of $w$ of order up to two vanish at $C$.  This violates Serrin's edge lemma (cf. \cite{Fr}), unless $w \equiv 0$ in $\Omega$, as desired.   \\

\noindent \emph{Case 2: The crest occurs at some point $(\lambda_0, \eta(\lambda_0))$, with $-L/2 < \lambda_0 < 0$, and the line $x = \lambda_0$ is normal to the free surface there.}  Put $\lambda_1 := 2\lambda_0+L/2$, $\lambda_2 := \lambda_0 + L/2$.  We wish to consider the domain that results from reflecting $D_{\lambda_1}$ over the line $x = \lambda_1$, and the set $D \setminus D_{\lambda_2}$ over the line $x = \lambda_2$.  More precisely, let us define
\[ \tilde{\eta}(x) := \left\{ \begin{array}{lc} 
\eta(2\lambda_0-x) &  \lambda_0 \leq x \leq \lambda_1, \\
\eta(2\lambda_2-x) & \lambda_1 \leq x \leq \lambda_2, \end{array} \right. \]
and 
\[ \Omega := \{ (x,y) \in \mathbb{R}^2 : \lambda_0 < x < \lambda_2, ~ -d<y< \tilde{\eta}(x) \}.\]
As the crest occurs at $(\lambda_0, \eta(\lambda_0))$, the profile must be non-increasing on the interval $x \in [\lambda_0, L/2]$.    We therefore observe that $\Omega \subset D$.  Let us further define, for all $(x,y) \in \Omega$,
\[ w(x,y) := \left\{ \begin{array}{lc}
\frac{\displaystyle \psi(x,y) -\psi(2\lambda_0 -x, y)}{\displaystyle \alpha(y)} & \lambda_0 \leq x \leq \lambda_1, \\
 & \\
\frac{\displaystyle \psi(x,y) - \psi(2\lambda_2 -x,y)}{\displaystyle\alpha(y)} & \lambda_1 \leq x \leq \lambda_2. \end{array} \right.\]
Suppose that we can prove $w \equiv 0$ in $\overline{\Omega}$.  Explicitly, this means $\psi(x,y) = \psi(2\lambda_0-x,y)$ in $\Omega$, for $\lambda_0 \leq x \leq \lambda_1$, and $\psi(x,y) = \psi(2\lambda_2-x,y)$ in $\Omega$ for $\lambda_1 \leq x \leq \lambda_2$.  Using the fact that $\psi_y < 0$,  we can differentiate the relation $\psi(x,\eta(x)) = 0$ to conclude $\eta(x) = \tilde{\eta}(x)$, for $x \in (\lambda_0, \lambda_2)$.  But this is impossible, as the free surface is assumed to be monotonic.  Thus $w \equiv 0$ in $\overline{\Omega}$ implies $\lambda_0 = 0$, and so we are in the first case.

In order to finish the proof of Case 2, we now argue that $w$ vanishes identically.  Let $C$ denote the crest point $(\lambda_0, \eta(\lambda_0))$ which lies at the upper left-hand corner of $\overline{\Omega}$.  Arguing as before, it is easy to show that $w \geq 0$ on the upper boundary and $w \equiv 0$ on the side and lower boundary of $\Omega$.  Likewise, simply computing as before, we show that under assumption \eqref{theorem1cond1}, $w$ is a supersolution in $\Omega$ with respect to the operator $\Delta + 2\alpha^{-1} \alpha_y \partial_y + c$, where $c \leq 0$.  Similarly, we check that all derivatives up to order two of $w$ vanish at $C$.  Then, since by definition $w(C) = 0$, Serrin's edge lemma produces a contradiction unless $w \equiv 0$ in $\Omega$.  Together with the argument of the previous paragraph, this completes the second case.  \\

\noindent \emph{Case 3:  $D_{\lambda_0}^R$ is internally tangent to the upper boundary of $D$ at some point $(\xi_1, \eta(\xi_1))$.}  Clearly, if $\lambda_0 = 0$, then the argument of {Case 1} applies and there is nothing additional to prove.  Without loss of generality, then, suppose $\lambda_0 < 0$.  Also, we note that $\xi$ must necessarily lie in some interval where the free surface is decreasing. This is a simple consequence of the fact that, left of the line $x = \lambda_0$, the boundary of $D_{\lambda_0}^R$ is negatively sloped.  We are therefore justified in defining $\lambda_1$, $\lambda_2$, $\tilde{\eta}$, $\Omega$ and $w$ as in {Case 2}.  

We know that $\psi = 0$ on the free surface and $\psi \geq 0$ in the interior of the fluid.  It follows that $w \geq 0$ on the upper boundary of $\Omega$.  As before, periodicity ensures that $\psi$ vanishes on the lateral boundaries of $\Omega$, by construction.  Finally, we know that $\psi = p_0$ on $y = -d$, hence $w$ vanishes identically on the lower boundary of $\Omega$.  

As in the previous cases, a simple calculation shows that $w$ is a supersolution in $\Omega$ with respect to a linear elliptic operator of the form $\Delta + \alpha^{-1} \alpha_y \partial_y + c$, where $c \leq 0$ when  \eqref{theorem1cond1} holds.    It follows that $w > 0$ in $\Omega$,  or $w \equiv 0$ in $\overline{\Omega}$.  Recall that we have already seen that the latter of these two possibilities coincides with {Case 1} and therefore implies symmetry.  

Seeking a contradiction, suppose instead that $w > 0$ in $\Omega$.     Let $C := (\xi_1, \eta(\xi_1))$ denote the contact point and note that it is in the interior of the upper boundary of $\Omega$.  We claim further that $w(C) = 0$.  To see why this is the case, observe that by definition of $\xi$, we have $\eta(\xi) = \tilde{\eta}(\xi)$.  Therefore, in light of the fact that $\psi = 0$ on the free surface, 
\[  \psi(\xi, \eta(2\lambda_0 - \xi)) = \psi(\xi, \eta(\xi)) = 0 = \psi(2\lambda_0 - \xi, \eta(2\lambda_0 -\xi)).\]
Thus $w$ attains a minimum at $C$.  Since $D_{\lambda_0}^R$ is internally tangent to $D$ at $C$, we must have that the interior sphere condition is satisfied there.  The hypotheses for Hopf's boundary lemma (cf. \cite{Fr}) now completely verified, we infer that
\[ \frac{\partial w}{\partial \nu}\bigg|_C < 0, \]
where $\nu$ is the normal vector at $C$.  

We seek to produce a contradiction by showing that, in fact, the normal derivative must vanish.  Towards that end let us consider $\xi_0 := 2\lambda_0 - \xi_1$, that is, the reflection of $\xi_1$ over the line $x = \lambda_0$.  By definition of $\xi_1$, we must have that $\eta(\xi_1) = \eta(\xi_0)$, and $\eta^\prime(\xi_1) = -\eta^\prime(\xi_0)$.  Differentiating the relation $\psi(x,\eta(x)) = 0$ and taking these observations into account, we arrive at the following identity
\be \frac{\psi_x}{\psi_y} \bigg|_{(\xi_0, \eta(\xi_0))} = - \frac{\psi_x}{\psi_y}\bigg|_C. \label{identity1} \ee
Here we again made use of the fact that $\psi_y = u - c< 0$.  Also, since $\rho$ is constant on the free surface, the nonlinear boundary condition \eqref{defQ} implies
\[ \left( |\nabla \psi|^2 + g\rho(-\psi) (\eta+d) \right)\bigg|_{(\xi_0, \eta(\xi_0))} = Q = \left( |\nabla \psi|^2 + g\rho(-\psi) (\eta+d) \right)\bigg|_{C},\]
 whence 
 \be |\nabla \psi (\xi_0,\eta(\xi_0))|^2 = |\nabla \psi(C) |^2. \label{identity2} \ee 
 From \eqref{identity1}-\eqref{identity2} we conclude further that 
 \[ \psi_x( \xi_0, \eta(\xi_0)) = -\psi_x(C), \qquad \psi_y(\xi_0, \eta(\xi_0)) = \psi_y(C).\]

 Let the normal vector at $C$ be $\nu = (\nu_1, \nu_2)$.  Then,
 \[ \frac{\partial w}{\partial \nu}\bigg|_C = \nu_1 \frac{\psi_x(C) + \psi_x(\xi_0, \eta(\xi_0))}{\alpha(\eta(\xi_1))} + \nu_2 \frac{\psi_y(C) - \psi_y(\xi_0, \eta(\xi_0))}{\alpha(\eta(\xi_1))} - \nu_2 \frac{\alpha_y(\eta(\xi_1))}{\alpha(\eta(\xi_1))^2} w(C), \]
 which is zero, by the above considerations.  We conclude that $w \equiv 0$ in $\overline{\Omega}$. This completes the proof that condition \eqref{theorem1cond1} implies symmetry.  
 
To prove \eqref{theorem1cond2} is sufficient, we note that the only use of \eqref{theorem1cond1} in the preceding argument was to ensure that a minimum principle holds for $w$.  If we take $\alpha \equiv 1$, then $w$ will be a solution of a linear elliptic operator of the form $\Delta + c$.  Since the ellipticity constant here is one,  {Lemma \ref{bnvtheorem1}} and  {Proposition \ref{bnvtheorem3}} together show that the requirement  
\be \sup_\Omega c^+ < \exp{\left(-\min\{L,~\eta_{\min}\}\right)}  \label{theorem1case2cond} \ee
implies $w$ has a minimum principle, and hence the wave is symmetric.  Condition \eqref{theorem1cond2} is chosen precisely so that \eqref{theorem1case2cond} holds.  

The last condition, \eqref{theorem1cond3}, will require more effort.  In particular, we shall work with the height equation formulation \eqref{heighteq} and make use of the operator $\mathcal{L}$.  In doing this we are closely following the path set forth by \cite{CEW}. 

Let $h$ be the height function for a steady periodic gravity water wave as in the statement.  If $h$ is flat there is nothing to proves, so without loss of generality we suppose that a crest occurs in the interval $[0,L/2)$ and a trough at $q = -L/2$.  This is permissible, as \eqref{heighteq} is invariant under the transformation $q \mapsto -q$.

For each $\lambda \in (-L/2, 0)$, $(q,p) \in [\lambda, 2\lambda + L/2] \times [p_0, 0]$, put
\[ q^{\lambda} := 2\lambda-q, \qquad w(q,p; ~ \lambda) := h(q,p) - h(q^\lambda, p).\]
Then, by construction, for $\lambda \in [-L/2,L/2]$, $(q,p) \in [\lambda, 2\lambda+L/2] \times [0,p_0]$,
\be w(\lambda, p; \lambda) = h(\lambda,p)-h(q^\lambda,p) = 0, \label{lambdalambdavanish} \ee
and 
\be w(q,p_0; \lambda) = h(q,p_0) - h(q^\lambda, p_0) = 0. \label{bottomvanish} \ee
Also we observe that
\be d\left(w(q,p; \lambda)\right) = \fint_T \left(h(q,0) -h(q^\lambda, 0)\right) dq = 0, \label{dvanish} \ee
owing to the periodicity of $h$.  

Fix $q$.  Then, by monotonicity of the profile, $w(q,0; \lambda) \geq 0$ for $\lambda$ near $-L/2$.  Put
\[ \lambda_0 := \sup \left\{ \lambda \in [-L/2, 0] : ~w(q,0; \lambda) \geq 0,~\textrm{for } q \in (\lambda, 2\lambda+ L/2)\right\}.\]
There are two cases to consider.\\

\noindent
\emph{Case 4}:  $\lambda_0 = 0.$  Denote $\Omega := (0,L/2) \times (p_0,0)$ and, for concision, fix $\lambda = \lambda_0$ for $w$.  Identity \eqref{bottomvanish} implies that $w \equiv 0$ on the set $\partial \Omega_b := \{ (q,p_0) : 0 < q < L/2 \}$.  Likewise,  by periodicity of $h$ we have $w(L/2, p) = h(L/2, p)-h(-L/2,p) = 0$, for all $p_0 \leq p \leq 0$.  Thus $w$ vanishes identically on the right boundary, $\partial\Omega_r := \{(L/2,p) : p_0 < p < 0 \}$.  Taking $\lambda$ and replacing it with $\lambda_0 = 0$ in \eqref{lambdalambdavanish}, we obtain the same result on the left boundary, $\partial\Omega_l := \{(0,p) : p_0 < p < 0\}$.  Finally, on the top, $\partial \Omega_t := \{ (q,0) : 0 < q < L/2\}$, we have $w \geq 0$ by the definition of $\lambda_0$. Combining then, we have $w \geq 0$ on $\partial\Omega$.  

Suppose for now that $w$ has a determined sign in some region $\Omega^\prime \subset \Omega$ whose boundary includes the corner point $(L/2,0)$ and which is blunt in the sense that the Serrin edge lemma holds there (see, for instance,  {Definition E3} of \cite{Fr}).  By \eqref{lambdalambdavanish} and periodicity, $w(L/2,0; \lambda_0) = w_p(L/2,0; \lambda_0) = w_q(L/2,0; \lambda_0) = 0$.   Using these facts we can differentiate \eqref{heightboundcondT} in $q$ and evaluate at the corner to
\[ 2 h_p h_{pq} \left(2g\rho h -Q\right) = 0, \qquad \textrm{at } (L/2,0), \]
which implies $h_{pq}$ --- and thus $w_{pq}$ --- vanishes at $(L/2,0)$.

As $w$ is a solution for $\mathcal{L}$ in $\Omega^\prime$ with determined sign, applying Serrin's edge lemma and the arguments of the previous paragraph we see that at least one of $w_{qq}$ and $w_{pp}$ must be nonvanishing at $(L/2,0)$.  This leads to a contradiction, as $w(q,p;\lambda_0) = h(q,p) - h(-q,p)$, and thus  
\[  \left. \begin{array}{lll} 
w_{qq}(L/2,0) &=& h_{qq}(L/2,0)-h_{qq}(-L/2,0) \\
w_{pp}(L/2,0) & = & h_{pp}(L/2,0)-h_{pp}(-L/2,0)  \end{array}\right\} = 0. \]
We conclude that no such region $\Omega^\prime$ can exist.  \\

\noindent
\emph{Case 5}:  $\lambda_0 \in (-L/2,0)$.  In this case, there also exists a first point $q_0 \in (\lambda_0, 2\lambda_0 + L/2]$ for which $w(q_0,0;\lambda_0) = 0$, and $w(q,0;\lambda_0) > 0$ for $\lambda_0 \leq q < q_0$  This will be attained at the point where the graphs of the functions $q \mapsto h(q,0)$, $q \mapsto h(q^{\lambda_0},0)$ are tangent to each other.  \\

First observe that, by definition of $\lambda_0$, we cannot have $w(q,0;\lambda_0) > 0$ for $q > \lambda_0$.  We may continuously extend $w$ by letting
\[ w(q,p;\lambda_0) := h(q,p) - h(q^{\lambda_0}+L,p), \qquad \textrm{for } (q,p) \in \left[2\lambda_0+\frac{L}{2},~\lambda_0+\frac{L}{2}\right] \times [p_0, 0].\]
Redefine 
\[ \Omega := \left(\lambda_0,~\lambda_0+\frac{L}{2}\right) \times (p_0, 0), \]
and, as in \cite{CEW}, periodicity implies $w \in C^2(\overline{\Omega})$.  Also, if $2\lambda+L/2$ lies to the left of the crest line, $w(q,0;\lambda) > 0$ holds for all $\lambda \leq q \leq 2\lambda+L/2$.  Thus $2\lambda_0 + L/2$ must be to the right of (or coinciding with) the crest line.  Monotonicity of the profile then ensures that $h$ is decreasing for $q \in (2\lambda_0+L/2,L/2)$.  This implies that on the top
\[ w(q,0; \lambda_0) \geq 0, \qquad \forall q \in \left[ \lambda_0,~\lambda_0+\frac{L}{2}\right].\] 
On the boundary, we therefore have the following sign information
\[ \left\{ \begin{array}{llll}
w(\lambda_0,p; \lambda_0)& = & w\left(\lambda_0+\frac{L}{2},p; \lambda_0\right) = 0, & \textrm{for } p \in [p_0, 0], \\
w(q,p_0; \lambda_0) & = & 0 \qquad \textrm{and} \qquad w(q,0; \lambda_0) \geq 0, & \textrm{for } q \in \left[\lambda_0, \lambda_0+\frac{L}{2}\right]. \end{array}\right.\] 

Suppose there exists a region $\Omega^{\prime\prime} \subset \Omega$ where $w \geq 0$ in $\Omega^{\prime\prime}$, $(q_0, 0) \in \partial\Omega^{\prime\prime}$ and $\partial\Omega^{\prime\prime}$ is smooth at $(q_0, 0)$.  Since the graphs of $q \mapsto h(q,0)$, $q \mapsto h(q^{\lambda_0},0)$ are tangent at the point $(q_0,0)$ this point, $w_q(q_0,0; \lambda_0) = 0$.  In particular, $h_q(q_0, 0) = -h_q(q^{\lambda_0},0)$.  Using this to evaluate \eqref{heightboundcondT} at $(q_0, 0)$ and $(q_0^{\lambda_0}, 0)$ we find $h_p(q_0,0) = h_p(q_0^{\lambda_0},0)$.  Thus $\nabla w(q_0,0; \lambda_0) = 0$.   However this contradicts the Hopf lemma (which is applicable, despite the adverse sign of the zeroth-order term of $\mathcal{L}$, because of the predetermined sign of $w$ in $\Omega^{\prime\prime}$, cf.  {Remark 2.16} of \cite{Fr}).  Thus no such region $\Omega^{\prime\prime}$ can exist. 

Proving both Case 4 and Case 5, therefore, involves showing that either $\Omega^\prime$ or $\Omega^{\prime\prime}$ must exist, unless $w \equiv 0$ in $\Omega$.   This can be done easily by reprising our arguments for \eqref{theorem1cond1}--\eqref{theorem1cond2}, provided we know $\mathcal{L}$ satisfies a maximum principle.  Unfortunately, this will not in general hold true.  However, if we let $\tilde{h}(q,p) := h(q^\lambda, p)$, then by  {Theorem \ref{bnvtheorem1}} we know that the maximum principle \emph{will} hold precisely when $\lambda_1(\mathcal{L}) > 0$.   {Lemma \ref{propLlemma}} then tells us that this will occur provided \eqref{eigenvaluecondition1} and \eqref{eigenvaluecondition2} hold.  Making this assumption, the rest of the proof of the theorem is immediate.  All that remains to show, therefore, is that \eqref{theorem1cond3} implies \eqref{eigenvaluecondition1}--\eqref{eigenvaluecondition2}. 

Let $\delta_1$, $\delta_2$ and $b$ be defined as in  {Lemma \ref{propLlemma}} taking $\tilde{h}(q,p) := h(q^{\lambda_0},p)$ for Case 1, or $h(q^{\lambda_0}+L,p)$ for Case 2.  Using the equation satisfied by $\tilde{h}$, we can estimate
\begin{align*} \sup_{\overline{R}} |\tilde{h}_{pp} | & \leq  \sup_{\overline{R}} \left|2 \tilde{h}_p \tilde{h}_q \tilde{h}_{qp} - \tilde{h}_p \tilde{h}_{qq} +g(\tilde{h}-d(\tilde{h})) \rho_p \tilde{h}_p^3 - \beta(-p) \tilde{h}_p^3 \right|  \\
& \leq 2 M^2 a_0^{-1} + M a_0^{-1}  +\left(\sup |\beta|+ g \eta_{\max} \sup |\rho_p| \right) a_0^{-3},\end{align*} 
where we take $M := \| h_q \|_{C^{1}(\overline{R})}$, which, in view of \eqref{defM2}, is precisely the same as in \eqref{defM}.  Then, from the above estimate find,
\begin{align*} b,~\delta_2 & \leq  4MA_0^2 +2M^2 A_0^3 + 3A_0 \left( \sup |\beta| + g \eta_{\max} \sup |\rho_p| \right) \nonumber \\ 
& \qquad + 2M^3 a_0^{-1} A_0^3 + M^2 a_0^{-1} A_0^3 + M a_0^{-3} A_0^3\left(\sup|\beta| + g\eta_{\max} \sup |\rho_p| \right). \end{align*}
The right-hand side, of course, we recognize as $\epsilon_2$.  Taking $\epsilon_1 := 3g \eta_{\max} \sup |\rho_p|$, \eqref{eigenvaluecondition1}--\eqref{eigenvaluecondition2} follow immediately from \eqref{theorem1cond3}.  This completes the theorem.  \end{proof}

\begin{remark} \label{dissusionofresult} From the above proof we can see quite clearly why providing criteria for the symmetry of stratified waves is dramatically more difficult than for the homogeneous case: it is \emph{precisely} the appearance of density variation that breaks the maximum principle structure.  Naturally, this leads us to wonder whether or not it is reasonable to expect that highly stratified waves are necessarily symmetric.  While we are not prepared to offer any conjectures on this question, it seems clear that, at the very least, making significant progress will require a new point of view.    
\end{remark}

\section{Acknowledgments}

The author owes a great debt to A. Constantin, M. Ehrnstr\"om, J. Escher, V. Hur, W. Strauss and E. Wahl\'en, who provided invaluable suggestions during the development of this work.  We are also grateful to the  many people who offered helpful commentary on an early version of this paper during the  ``Wave Motion'' and ``Low Eigenvalues of Laplace and Schr\"odinger Operator'' workshops, held concurrently in February, 2009 at Oberwolfach. \\

\end{document}